\documentclass[12pt]{article}
 \usepackage{graphicx} \usepackage{epsfig}
 \usepackage{bm} \usepackage{amsmath} \usepackage{amssymb}

\begin{document}

\begin{center}
 \textbf{\huge{ Renormgroup symmetry for \\
 solution functionals}}\\  \bigskip

{\normalsize D.~V.~Shirkov $^{a\,,\,b}$,
            V.~F.~Kovalev  $^c$,} 

{\scriptsize $^{a}$ Joint Institute for Nuclear Research, Dubna}
{\scriptsize $^{b}$ M.V.Lomonosov Moscow State University} \\
{\scriptsize $^{c}$ Institute for Mathematical Modelling RAS,
                    Moscow } 
\end{center} 

\textit{\small The paper contains generalization of the
renormgroup algorithm for boundary value problems of mathematical
physics and related concept of the renormgroup symmetry,
formulated earlier by authors with reference to models based on
differential equations. These algorithm and symmetry are
formulated now for models  with non local (integral) equations. We
discuss in detail  and illustrate by examples applications of the
generalized algorithm to models with not local terms which appear
as linear functionals of the solution.}\vspace{-2mm}


\section{Introduction}

The concept of RenormGroup Symmetry (RGS) appeared in
mathematical physics in the beginning of
90s~\cite{kov-tmf-89, shkkp-rg-91} (see, also reviews
\cite{kov-jmp-98,ksh-phr-01}) being borrowed from theoretical physics.
In turn, the concept of the Renormalization Group (RG) for the first time
has appeared in the most complex branch of last --- in the
Quantum Field Theory (QFT). A presence of a group structure (Lie
groups of transformations) in the QFT calculation results was
discovered in the beginning of 50s by Stueckelberg and Petermann
\cite{stp-hpa-51} (see, also \cite{gml-phr-54,bsh-dan-55a}).\par

This structure and exact symmetry of a solution underlying have
been then used by N.N.Bogoliubov for creation of regular method
for improving of an approximate solution of QFT problems -- the
Renormalization Group Method (=RGM) \cite{bsh-dan-55b} (see
also~\cite{bsh-nc-56, bsh-jetf-56,bsh-bk-80}). This method
contains elements of the theory of continuous group
transformations (theory of Sophus Lie). Improvement of
approximation properties appears to be most essential in a
vicinity of the solution singularity.
 \par

Turn now to the fact that in the case of problems described by
complicated equations --- as, e.g., in the transfer theory
(integro-differential Boltzmann equation) or in the quantum
field theory (an infinite chain of engaged integro-differential
Dyson-Schwinger equations) --- only some components of solution
or their integrated characteristics obey enough simple symmetry.
So, in QFT the central object in RG transformations is the so
called ``function of an invariant charge'' $\bar{\alpha}$ (or
``running coupling constant"), representing specific product of
Lorentz-invariant amplitudes of propagators $d_i\,,$ vertices
$\Gamma_k\,$ and the expansion parameter $\alpha\,$
 \[
 \bar{\alpha} = \alpha\,\Gamma^2 \prod \limits_{i} d_i \,.
 \]
At the same time, Schwinger-Dyson equations include only
functions $d_i\,\Gamma_k\,$ and the coupling constant
$\alpha\,,$ separately, but not their product $\bar{\alpha}\,.$
In the one-velocity plane transfer problem the RG-invariance
property is related to the asymptotics of ``density of
particles, moving deep into the medium'' $n_{+}(x)\,, \ x \to
\infty\,$ not entering the Boltzmann equation\footnote{And
representable as the integral $\int\limits_0^1 n(x,\vartheta)\,
{\rm d}\cos\vartheta\,$  of the solution of kinetic equation
$n(x,\vartheta)\,$.}.
 \par

The renormalization group concept was transferred to
mathematical physics \cite{kov-tmf-89,shkkp-rg-91} with the same
pragmatic goal, as in QFT, in mind --- ``improvement" of the
solution behaviour in vicinity of a singularity. Remind that
proliferation of the RG method from QFT to others fields of
theoretical physics (the theory of critical phenomena, physics
of polymers and so on) has caused various and sometimes
essentially different (in comparison with initial) forms of
realization of RG ideology (see, for example, the review
\cite{shr-umn-94}). In application to problems of mathematical
physics there appeared different variants of formulations of RG
method \cite{gol-pre-96}-\cite{bri-kup-90s}.
\par

For the boundary value problems (BVPs) of mathematical physics
using DE, we developed (see, e.g.,
\cite{kov-tmf-89,shkkp-rg-91}, and also recent reviews
\cite[p.232-249]{ksh-phr-01}, \cite{kov-rg-02}) the RG
algorithm, which is quite distinct from earlier ones. To this
difference became obvious, we remind, that in a basis of the RG
Method, as it was initially formulated by N.N.Bogoliubov and one
of authors \cite{bsh-dan-55b} for QFT problems, lays the use of
an \textit{exact} group property of a solution. One of the
well-known formulations of this property is the functional
equation (representing only a group composition law) for the
invariant charge in QFT. In every concrete case, revealing of
similar symmetry (i.e., of group property) for the solution
demands a special,  usually non-standard, analysis (see, for
example, discussion in the papers
\cite{shr-ijmp-88,shr-umn-94,gol-pre-96}), that is an
algorithmic drawback of the RG technique. \par

Coming back to mathematical physics (MP) we note, that here we
usually deal with the problems based on systems of DEs the
symmetry of which can be found by a regular way with the help of
Lie group analysis. In problems of MP this feature appeared as
decisive in creating ``RG-algorithm" which has united RG ideology
of QFT with a regular way of symmetry construction for BVP
solutions. Due to this algorithm also there arised  the concept of
``renormgroup symmetry" for solutions of BVP: these symmetries
result from  calculation procedure similar to that used in the
modern group analysis.
 \par

At the initial stage \cite{kov-tmf-89,shkkp-rg-91}, application
of RG algorithm was mainly limited to problems based on DEs
though formally this algorithm can be used in any problem, for
which it is possible to specify a regular way of calculation
symmetries for basic equations. Hence, transition to such
objects, which up to a recent time were not a subject of the
group analysis, in particular, to the integral and
integro-differential equations, essentially expands the area of
RGS applications.\par

In just mentioned cases integral relations form a skeleton of a
problem. They, however, can appear as some independent objects
for application of RGS, constructed for solutions of DE.
Frequently, of physical interest is not the solution itself in
all range of change of variables and parameters, but rather its
certain integral characteristic, a solution functional. This
characteristic can appear, for example, in result of averaging
(integrating) over one of independent variables\footnote{see,
for example, previous footnote.} or when passing to new integral
representation, e.g., to Fourier representation. In this case,
RG algorithm can be applied not for improving of a particular
solution with the subsequent calculation of its integral
characteristic, but directly for improvement of the solution
functional for an approximate solution. These motives gave a
stimulus for expanding of the RG algorithm to models with
non-local (integral) equations. \par

The contents of the paper is structured as follows. In section 2
an introductory example to RGS algorithm in mathematical physics
is presented, illustrated by a solution of a simple BVP. In
section 3, generalization of RG algorithm, developed in
application to BVP for DE \cite{kov-jmp-98} (and reviewed in
\cite{ksh-phr-01} and \cite{kov-rg-02}), is formulated for
models with non-local equations. In section 4, the review of
recent results received on the basis of the modified RG
algorithm is presented, and also efficiency of a method in
application to some already known solutions is shown. Some
generalities, uniting concepts of RG symmetry, the functional
self-similarity, are considered in Appendix.

\section{The introductory example to the RGS algorithm}  

Generally, the RG can be defined as a continuous one-parameter
group of specific transformations of a partial solution (or
solution characteristic) of a problem, a solution that is fixed
by boundary conditions. The RG transformation involves boundary
condition parameters and corresponds to some change in a way of
imposing this condition. \par

For illustration, consider transformations $T_a$,
  \begin{equation}
 \begin{aligned}
 {\bar{x}}^i = f^i(x,a) \, , \quad
 f^i(x,a_0) = x^i \,, \quad i=1, \ldots , n \,,
 \end{aligned}
 \label{tranformations}
  \end{equation}
depending on a real parameter $a$, where $x\in \mathbb{R}^n$. A
set $G$ of these transformations form a one-parameter local
group if the functions $f^i(x,a)$ satisfy the composition rule
$T_b T_a= T_{\phi(a,b)}$,
 \begin{equation}
 \begin{aligned}
 & f^i(f^i(x,a),b)= f^i(x,\phi(a,b))\, , \quad  \phi(a,a_0)= a \,
 , \ \phi(a_0,b)= b \, ,
 \end{aligned}
 \label{composionrule}
 \end{equation}
that can be transformed to the simplest form with $\phi(a,b)= a
+ b $ and with $a_0=0$ in (\ref{tranformations}).

For a given solution of some physical problem renormgroup
transformations in the simplest case are defined as
transformations of (\ref{tranformations}) type, i.e. as \textit{
simultaneous one-parameter group transformations $R_t$ of two
variables}, say $x$ and $g$,
 \begin{equation}
 \begin{aligned}
 x \to x^{\prime} = x / t \,, \quad
 g \to g^{\prime} = {\bar{g}}(t,g)
 \,,
 \end{aligned}
 \label{rg-simple}
  \end{equation}
the first being the scaling of a coordinate $x$ (or reference
point) and the second -- a more complicated functional
transformation of the solution characteristic $g$. Hence, the RG
transformation corresponds to a change in the parametrization
for the \textit{same} solution, while the equation
(\ref{composionrule}) for the function $\bar{g}$ has the form
 \begin{equation}
 \begin{aligned}
 & {\bar{g}}(x,g)= {\bar{g}} \left( x/t,{\bar{g}}(t,g) \right) \,,
 \quad {\bar{g}}(1,g) = g \,,
 \end{aligned}
 \label{funceq}
 \end{equation}
and guarantees the group property $R_{\tau t}=R_{\tau}R_{t}$
fulfillment for transformations (\ref{rg-simple}). These are
just the RG functional equations and transformations for a
massless QFT model with one coupling constant
\cite{bsh-dan-55b}. In that case $x=Q^2/\mu^2$ is the ratio of a
four-momentum $Q$ squared to a ``normalization'' momentum $\mu$
squared and $g$ is a coupling constant, while $\bar{g}$ is the
so-called effective coupling.
\par

Geometrically, transformations (\ref{tranformations}) mean that
any point $x \in \mathbb{R}^n$ is carried by this
transformations into the point $\bar{x}$ whose locus is a
continuous curve passing through $x$, known as of a path curve
of the group $G$. The group property (\ref{composionrule}) means
that any point of a path curve is carried by $G$ into points of
the same curve. The locus of the images $T_a(x)$ is also termed
the $G$-orbit of the point $x$. The correspondence between
transformations (\ref{tranformations}) and (\ref{rg-simple})
means that for RG transformations a curve in the $\{x,g\}$ plane
that defines the solution of a physical problem is the path
curve of the renormgroup $R_t$. In other words, the solution of
the problem coincides with the $R_t$-orbit of a boundary
manifold -- the point $\{x=x_0,g=g_0\}$. Upon the RG
transformations the reference (boundary) point $\{x_0,g_0\}$ is
shifted to some other value $\{x_1,g_1\}$, while the solution
remains unaltered, i.e. the solution curve is the invariant
manifold of the group $R_t$ (like the invariant charge in QFT
\cite{bsh-bk-80}).
\par

Hence, the general problem of searching for the RG
transformations may be reformulated as follows: the solution of
the physical problem should coincide with the orbit of the
renormalization group.
\par

In mathematical physics a solution of a physical problem usually
appears as a solution of some BVP. Then the corresponding RG
transformations may be obtained from the symmetry group related
to this BVP, provided the boundary condition is also involved in
group transformations. The key point here is the fact that the
corresponding symmetry group are calculated using the regular
algorithms of modern group analysis, provided the basic
mathematical model is formulated in terms of differential (or
integro-differential) equations.
\par

Let this model be given by a system of $k-$th order differential
equations, identified with its frame,
 \begin{equation}
 \begin{aligned}
 & F_{\sigma}(x,u,u_{(1)},\ldots , u_{(k)})=0\,, & \quad
 & \sigma =1,\ldots , s \,.
 \end{aligned}
  \label{rgmloc-int} \end{equation}
In the paper we use the terminology of differential algebra and
notations for variables accepted in the group analysis
\cite{ibr-spr-94}:
 \begin{equation}
 \label{var-def}
  x = \{ x^i \}, \quad
  u = \{u^\alpha\}, \quad
  u_{(1)} = \{u^\alpha_i\}, \quad
  u_{(2)} = \{u^\alpha_{ij}\}, \ldots ,
 \end{equation}
where $\alpha = 1,\ldots , m; \ i,j,\ldots = 1,\ldots , n$.
Variables $x$ and $u$ are referred to as independent variables
and \textit{differential variables}, respectively, having
consecutive derivatives $u_{(1)}$, $u_{(2)}$, \ldots etc.
Differential variables are related by a system of equations
 \begin{equation} \label{rel-def}
 \begin{aligned} &
      u^{\alpha}_{i} = D_i \left( u^{\alpha} \right) \,, \quad
      u^{\alpha}_{ij} = D_j \left( u^{\alpha}_{i} \right) =
                        D_j D_i \left( u^{\alpha} \right)\,, \ldots
 \end{aligned}   \end{equation}
via the operator of the total differentiation
 \begin{equation} \label{totaldif}
  D_i = \frac{\partial}{\partial z^i} + u^{\alpha}_{i}
  \frac{\partial}{\partial u^{\alpha}} + u^{\alpha}_{ij}
  \frac{\partial}{\partial u^{\alpha}_{j}} +  \ldots \, .
 \end{equation}
Locally analytical function of variables (\ref{var-def}), for
example, the function $F(x, u, u_{(1)}, \ldots, u_{(k)}) $ with
the highest order derivative $k$ refers to as differential
function of the $k$-th order, and a set of all such functions
with any values of $k\,$ forms the space of differential
functions ${\cal {A}} [x, u] $. Any function $F \in {\cal{A}}
[x, u] $ gives rise to a differential manifold $[F]$, determined
by an infinite system of equations
 \begin{equation} \label{man-def}
    [F]: \quad F=0\,, \quad D_i F =0\,,
         \quad D_i D_j F = 0 \,, \ldots \, .
 \end{equation}
The manifold $[F]$ is called the frame of the $k$-th order
partial differential equation
 \begin{equation} \label{deq-frame}
    F \left( x,u, \frac{\partial u}{\partial x} , \ldots ,
    \frac{\partial^k u}{\partial x^k} \right)=0 \, .
 \end{equation}
According with the \textbf{\textit{definition}} a system of
$s$-th order differential equations is said to be invariant
under a group $G$ if the frame of the system is an invariant
manifold for the extension of the group $G$ to the $s$-th order
derivatives \cite[p.209]{ibr-bk-96}. When utilizing an
infinitesimal group generator
  \begin{equation} \label{oper}
  X=\xi^{i}\partial_{x^i} + \eta^{\alpha} \partial_{u^{\alpha}}\,,
  \quad \xi^{i}\,, \eta^{\alpha} \in \cal{A}\,,
  \end{equation}
with coordinates $ \xi^{i} \, , \ \eta^{\alpha}$, which are
functions of group variables $\{x^i, u^{\alpha} \}$, this
definition leads to the invariance criterion in the following
form
 \begin{equation}  \label{deteq}
  {X_{(k)}\, F_{\sigma}}_{\Big\vert {[F_{\sigma}]}} = 0 \,, \quad
 \sigma=1,\ldots{},s\,\,\,,
 \end{equation}
where $X_{(k)}$ denotes $X$ extended to all derivatives,
involved in $F_{\sigma}$ and the symbol $\vert_{[F]}$ means
evaluated on the frame (\ref{man-def}). Solving a system of
linear homogeneous partial differential equations (known as
\textit{the determining equations}) for coordinates $ \xi^{i} \,
\ \eta^{\alpha} $  gives a set of infinitesimal operators
(\ref{oper}) (or group generators) which correspond to the
admitted vector field of the symmetry group $G$ and form a Lie
algebra $L$.
\par

Let the Lie group $G$ with the generator
  \begin{equation} \label{oper-first}
  X=\xi^{t} \partial_{t} + \xi^{x} \partial_{x} + \eta \partial_{y}\,,
  \end{equation}
be defined for the system of the first order partial
differential equations
 \begin{equation} \label{first-eq}
    y_t = F \left( t,x,y,y_x \right) \, .
 \end{equation}
The typical BVP for (\ref{first-eq}) is the Cauchy problem with
the boundary manifold defined by
 \begin{equation} \label{first-boundary}
    t=0 \,, \qquad y= \psi (x) \, .
 \end{equation}
The solution of this Cauchy problem is the $G$-invariant
solution, iff for any generator (\ref{oper-first}) the function
$\psi$ obeys the Equation \cite[\S29]{ovs-bk-82}:
  \begin{equation} \label{invar-first}
 \eta (0,x,\psi) - \xi^{x}(0,x,\psi) \partial_{x} \psi
 - \xi^{t} (0,x,\psi) F (0,x,\psi,\partial_{x} \psi) = 0 \,.
  \end{equation}
The solution of the Cauchy problem
(\ref{first-eq})--(\ref{first-boundary}) coincides with orbit of
the group $G$ and the boundary manifold is partially invariant
manifold of the group $G$ with the defect $\delta=1$.
\par

This example gives an instructive idea of constructing
generators of renormgroup symmetries. The milestones here are:
a) considering the BVP in the extended space of group variables
that involve parameters of boundary conditions in group
transformations, b) calculating the admitted group using the
infinitesimal approach, c) checking the invariance condition
akin to (\ref{invar-first}) with the goal to find the symmetry
group with the orbit which coincides with the BVP solution and
d) utilizing the RG symmetry to find the improved (renormalized)
solution of the BVP.
\par

The full algorithm was described in details in our previous
publications \cite{kov-jmp-98,ksh-phr-01} and will be also
touched upon in the next section while here we will only give a
general grasp at the problem using a trivial example, the BVP
for the Hopf equation
 \begin{equation}
 v_z + v v_x = 0\,, \quad v(0,x)= \epsilon U(x)\,,
 \label{hopf}
 \end{equation}
where $U$ is the invertible function of $x$. Introducing
$u=\epsilon v$ we insert the boundary amplitude directly in the
input equation
 \begin{equation}
 u_z + \epsilon u u_x = 0\,, \quad u(0,x)= U(x)\,.
 \label{hopf-eq}
 \end{equation}
For small values of $\epsilon z \ll 1$, i.e. near the boundary,
$z\to0$, or for small amplitude at the boundary, $\epsilon \to
0$, a perturbation theory solution to (\ref{hopf-eq}) has the
form of a truncated power series in $\epsilon z$,
 \begin{equation}
 u = U - \epsilon z U U_x + O\left( (\epsilon z)^2 \right) \,.
 \label{hopf-pt} \end{equation}
It is obvious that this solution is invalid for large distances
from the boundary, when $ \epsilon z U_x \backsimeq 1 $. The
renormgroup symmetry gives the root to improving the
perturbation theory result and restore the correct structure of
the BVP solution.
\par

With the goal to obtain this symmetry we extend the list of
variables, involved  in group transformations, adding the
parameter $\epsilon$ to the list of independent variables. Then
we calculate the admitted symmetry group $G$, with the generator
  \begin{equation} \label{oper-hopf}
  X=\xi^{z} \partial_{z} + \xi^{x} \partial_{x}
    + \xi^{\epsilon} \partial_{\epsilon}
    + \eta \partial_{u}\,,
  \end{equation}
using the classical Lie calculational algorithm (see, e.g.
\cite{ibr-bk-96}) that employs the infinitesimal criterion
(\ref{deteq}). The solution of determining equations gives
coordinates of the generator (\ref{oper-hopf}),
  \begin{equation} \label{coord-hopf}
 \xi^{z} = \psi^1 \,, \quad
 \xi^{x} = \epsilon u \psi^1 + \psi^2 + x(\psi^3 + \psi^4)\,, \quad
 \xi^{\epsilon} = \epsilon \psi^4 \,, \quad
 \eta = u \psi^3 \,,
  \end{equation}
where $\psi^i\,,\ i=2,3,4$, are arbitrary functions of
$\epsilon, u,x-\epsilon u z$ and $\psi^1$ is an arbitrary
function of all group variables. These formulas define an
infinite-dimensional Lie algebra with four
generators\footnote{In case when the amplitude $\epsilon$ is not
involved in transformations we have only three generators (see,
e.g. \cite[p.222]{ibr-spr-94}).}
  \begin{equation} \label{gen-hopf}
  X_1=\psi^{1} \left( \partial_{z} +  \epsilon u \partial_{x}
  \right) \,, \
  X_2=\psi^{2} \partial_{x} \,, \
  X_3=\psi^{3} \left(x \partial_{x} +  u \partial_{u} \right) \,, \
  X_4=\psi^{4} \left(\epsilon \partial_{\epsilon}
                  +  x \partial_{x} \right) \,.
  \end{equation}
Suppose we are given a particular solution of the BVP
(\ref{hopf-eq}), $u-W(z,x,\epsilon)=0$, which defines an
invariant manifold of the group (\ref{oper-hopf}),
(\ref{coord-hopf}). The corresponding invariance condition,
evaluated on the frame (\ref{hopf-eq}), looks similar to
(\ref{deteq}),
 \begin{equation} \label{invarcond-hopf}
 (W-xW_x)\psi^3 - W_x \psi^2 - (\epsilon W_\epsilon+xW_x)\psi^4 = 0 \,.
  \end{equation}
This equation is valid for all $z$, hence it remains valid for
$z=0$, when $W$ is replaced by $U(x)$. In this limit, $z\to 0$,
the condition (\ref{invarcond-hopf}) gives a relationship
between $\psi^i\,,\ i=2,3,4,$ (no restrictions are imposed on
$\psi^1$), that can be easily prolonged on $z \neq 0$,
 \begin{equation} \label{restrict-hopf}
 \psi^2 = - \chi ( \psi^3 + \psi^4 ) + ( u/U_{\chi}) \psi^3 \,, \quad
  \chi = x - \epsilon u z \,,
  \end{equation}
where the derivative $U_{\chi}$ should be expressed either in
terms of $\chi$ or $u$ in account of the boundary condition.
Inserting (\ref{restrict-hopf}) in (\ref{coord-hopf}) we get the
group of a smaller dimension with the generators
  \begin{equation} \label{rg-hopf}
  R_1=\psi^{1} \left( \partial_{z} +  \epsilon u \partial_{x} \right) \,, \
  R_2= u \psi^{3} \left[ \left( \epsilon z +1/U_{\chi} \right) \partial_{x}
       + \partial_{u} \right]  \,, \
  R_3= \epsilon \psi^{4} \left(  z u  \partial_{x}
                + \partial_{\epsilon} \right) \,.
  \end{equation}
The above procedure, that transforms (\ref{gen-hopf}) to
(\ref{rg-hopf}) we refer to as \textit{the restriction of the
group} (\ref{oper-hopf}) \textit{on a particular solution}.
\par

The solution of the BVP defines a manifold, which appears to be
invariant for any generator $R_i$ just by a method of
construction, hence (\ref{rg-hopf}) defines the desired RG
symmetries. This means that the solution of the BVP can be
constructed using any generator of the RG algebra
(\ref{rg-hopf}), say, the generator $R_3$. Without loss of
generality we choose $\epsilon \psi^{4}=1$ and obtain the
following finite RG transformations ($a$ is a group parameter)
 \begin{equation}
 \begin{aligned}
  x^{\prime} = x + a z u \,, \quad
  \epsilon^{\prime} = \epsilon + a \,, \quad
  z^{\prime} = z \,, \quad   u^{\prime} = u \,.
 \end{aligned}
 \label{trans-hopf}
  \end{equation}
Here $z$ and $u$ are invariants of RG transformations, while
transformations of $\epsilon$ and $x$ are translations, which in
case of $x$ also depend upon $z$ and $u$. For $\epsilon = 0$ we
have, in view of (\ref{hopf-pt}), $x = H(u)$, where $H(u)$ is
defined as the function inverse to $U(x)$. Then eliminating $a$,
$z$ and $u$ from (\ref{trans-hopf}) and omitting primes over the
variables, we get the desired solution of the BVP
(\ref{hopf-eq}) in the non-explicit form
 \begin{equation}
 \begin{aligned}
  x - \epsilon z u = H(u) \,.
 \end{aligned}
 \label{solution-hopf}
  \end{equation}
This in fact is the improved PT solution (\ref{hopf-pt}), which
is valid not only for small $\epsilon \ll 1$, provided the
dependence (\ref{solution-hopf}) can be resolved in a unique
way. \par

The peculiarity of the procedure of constructing RG symmetries
is the multi-choice first step that depends on the way in which
boundary conditions are formulated  and the form in which the
admitted symmetry group is calculated. For example, instead of
calculating the Lie point symmetry group we may consider the
Lie-B\"acklund symmetries \cite{ibr-bk-85} with the canonical
generator $R = \kappa\partial_u$, where $\kappa$ depends not
only on $z,x,\epsilon,u$, but on higher order derivatives of $u$
as well. We may look for $\kappa$ in the form of a power series
in $\epsilon$, and the invariance condition
(\ref{invarcond-hopf}) is formulated as vanishing of $\kappa$ at
$z \to 0$. Depending on the choice of the zero-order term
representation we get either infinite or a truncated power
series for $\kappa$, say, a linear in $\epsilon$ form,
  \begin{equation} \label{cangen-hopf}
 R = \kappa \partial_{u} \,,\quad
 \kappa = 1- \frac{u_x}{U_x(u)} - \epsilon z u_x \,.
  \end{equation}
This RG generator (\ref{cangen-hopf}) is equivalent to the Lie
point generator $R_2$ from (\ref{rg-hopf}), thus giving the same
result.
\par

Another possibility of calculating RG symmetries for BVP
(\ref{hopf-eq}) is offered by taking into account some
additional differential constraints which are consistent with
the boundary conditions and input equations. For example, if the
boundary condition in (\ref{hopf-eq}) is linear in its argument,
$U(x)=-x$, the differential constraint may be taken as
$u_{xx}=0$; this equality reflects the invariance of the
original equation with respect to the second-order
Lie-B\"acklund symmetry group. Calculating the Lie point
symmetry group for the joint system of this constraint and the
Hopf equation gives another way to finding RG symmetries for the
BVP (\ref{hopf-eq}).
\par

The above example demonstrates the milestones of the RGS method
in mathematical physics. The details of the general approach
will be discussed in the next section. Here we just point to the
fact that in order to construct RGS we employ the symmetry group
calculated in a regular way using the modern group analysis
technique. This group is considered in the extended space of
variables that includes the parameters and boundary data
involved in group transformations. The invariance of the
perturbative solution is used to find the particular RGS
generator that leaves the particular BVP solution unaltered.
Then the utilization of the finite group transformations restore
the desired structure of the BVP solution.

\section{The scheme of RG algorithm for non-local problems}  

A procedure of construction and use of RGS with reference to BVP
for DE is described in details in our previous works, for
example, in reviews \cite{kov-jmp-98,ksh-phr-01} and illustrated
in the previous section. Nevertheless, for connectivity of a
statement, the basic stages of the scheme are briefly depicted
below with accent on those changes, which it is necessary to
introduce in RG algorithm in order to make it applicable to
non-local problems.
\par

The starting point in these problems is the system of $\nu\geq 1\,
$ integro-differential (including differential and integral)
equations for functions $u =\{u^{\alpha} \}$, $\alpha= 1,\ldots,m
$ of variables $x = \{x^{i}\} $, $i = 1, \ldots, n $,
\begin{equation} \label{mathmodel}
 \begin{aligned}
& [E]: \qquad E_{\nu}(u(x))=0\,,
\end{aligned} \end{equation}
with the non-local terms depending  upon integrals of these
functions, and supplemented by appropriate boundary or initial
conditions. We assume, that we know some approximate solution,
$u^{\alpha} =U^{\alpha} $, for example, represented by a
truncated power series in powers of a small parameter or in
powers of a small deviation from a boundary of an area, where
the solution is known. \par

\begin{figure}[!t]
 \includegraphics[width=0.9\textwidth]{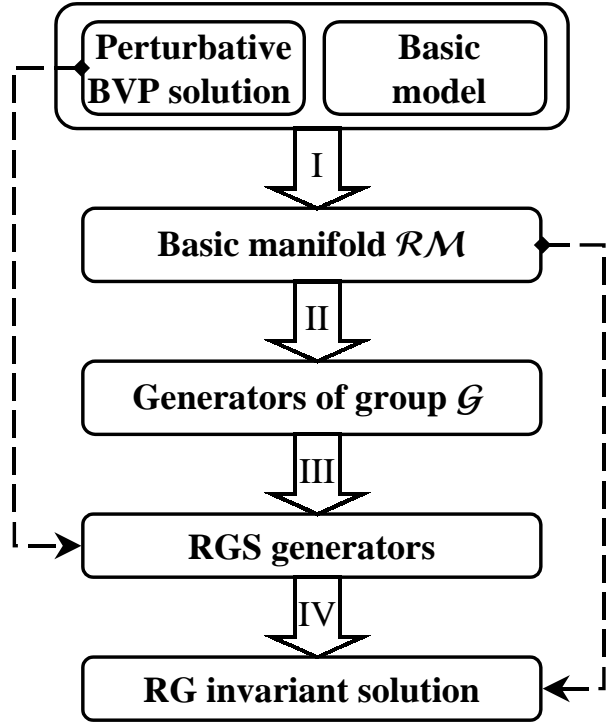}
 \caption{General scheme for realization of RG algorithm.}
 \label{fig1}
 \end{figure}

Then, conditionally, the scheme of realization of RG algorithm
can be expressed as a sequence of four steps, submitted by  the
scheme on the Figure~\ref{fig1},
\\
 \textbf{(I)} construction of basic \textit {manifold}, \\
 \textbf{(II)} calculation of the admitted \textit {symmetry group} and \\
 \textbf{(III)} its \textit{restriction} on a particular BVP
               solution leading to revealing of RGS, and also \\
 \textbf{(IV)} searching for \textit{an analytical solution} that is
              adequate to this RGS. \par

\medskip

 \par

\subsection{Construction of RG manifold}{\label{RGM}}  

The initial problem is the construction of RGS and appropriate
transformations that also touch upon parameters of particular
solutions. Therefore the purpose of \textbf{a first step (I)}
consists of involving in group transformations in this or that
way the parameters entering both in the equations of a problem,
and in boundary conditions on which this partial solution
depends. This purpose is achieved by construction of a special
manifold $\cal{RM}$ which we believe is given as a system of $s$
DE of the $k$-th order and $q$ non-local relations,
 \begin{alignat}{2}
 & F_{\sigma}(z,u,u_{(1)},\ldots , u_{(k)})=0\,, & \quad
 & \sigma =1,\ldots , s \,, \label{rgmloc} \\
 & F_{\sigma}(z,u,u_{(1)}, \ldots , u_{(r)}, J(u)) = 0 \,, & \quad
 & \sigma = 1 + s , \ldots , q + s \,.\label{rgmnonloc}
 \end{alignat}
Non-local variables $J(u)$ included in (\ref{rgmnonloc}) are
introduced by integrated operations
 \begin{equation}  \label{nonlocalvar}
 J(u) = \int {\cal F}(u(z)) {\rm{d}} z\,.
 \end{equation}
A presence of relations (\ref{rgmnonloc}) characterizes the
basic difference of $\cal {RM} $ for non-local problems from the
case of BVP for DE, for which the manifold $\cal{RM}$ is
differential.
\par

Let's note that generally $\cal{RM}$ \textit {does not coincide}
with a system of initial equations. Only on occasion, which it
will be told about below, and at the additional clauses
expanding and specifying the list of variables and the
parameters entering in RG transfor\-ma\-tions, it is possible to
establish conformity between (\ref{mathmodel}) and
(\ref{rgmloc})--(\ref{rgmnonloc}). However it is not possible to
execute a first step of the algorithm for any boundary problems
simply treating $[E]$ as $\cal{RM}$, and the concrete form of a
realization of the first step depends both on a form of the
initial equations (\ref{mathmodel}), and on the form in which
boundary conditions are presented. Formulated earlier for models
with DE the general approach to construction of $\cal{RM}$
remains also valid for non-local problems. In this sense several
specified earlier (see, for example, \cite{ksh-phr-01}) and
illustrated in section I ways of constructing this manifold are
possible. These ways do not exhaust all opportunities to
construction of $\cal{RM}$, rather they emphasize a variety of
approaches to realization of a first step depending on a
character of a problem under consideration. A choice of a
concrete realization more often is dictated both by a form of
the Eq.~(\ref{mathmodel}) and boundary conditions to them, and a
type of the approximate solution. Such multi-variant situation
is inherent only in the first step of the algorithm and is aimed
on covering the widest variety of the problems, analyzed by this
method. Already the following step of the scheme is carried out
in frameworks of the well-developed group-theoretical methods.

\subsection{Calculation of the transformation group \label{GROUP}}

\textbf{The next} step \textbf{(II)} consists in calculation the
most wide admitted symmetry group $\cal{G}$ for
Eqs.~(\ref{rgmloc})-(\ref {rgmnonloc}). Here the essential
change of RG algorithm is required in comparison with its
realization for differential manifold $\cal{RM}$. Really, in
application to $\cal{RM}$, defined only by a system of DE
(\ref{rgmloc}), the question was about a local group of
transformations in space of differential functions $ \cal{A} $,
at which  system (\ref{rgmloc}) remains unchanged. \par \medskip

At transition to manifold $\cal{RM}$ which is set by the system
of Eqs.~(\ref{rgmloc})-(\ref{rgmnonloc}), classical Lie
algorithm, using the infinitesimal approach, appears
inapplicable. The basic obstacle here is that ${\cal{RM}}$ in
this case is not determined \textit{locally} in space of
differential functions, therefore the basic advantage of Lie
computational algorithm, namely, representation of DetEq as the
over-determined system of equations is not realized here. Also
in frameworks of the classical group analysis the procedure of
prolongation of the group operator of point transformations on
non-local variables is not defined. Probable ways to overcome
these complexities while performing the second step of RG
algorithm are specified below.\par

At modification of the RG algorithm we lean on the direct method
of calculation of symmetries which was advanced in
\cite{mel-dan-87}--\cite{kov-pjetf-92} and used for finding
symmetries for the Boltzmann kinetic equation, the equations of
motion of viscous-elastic media and Vlasov-Maxwell equations in
the kinetic theory of plasma. This method is based on a
generalization of a group of symmetry, the so-called
Lie-B\"acklund symmetry group (terms ``higher'' or
``generalized'' sym\-metry are also in use), defined by the
generator of the form (\ref{oper}) prolonged on all higher-order
derivatives,
\begin{equation} \label{lb-oper}
 \begin{aligned} &
  X=\xi^{i}\partial_{z^i} + \eta^{\alpha} \partial_{u^{\alpha}}
 +\zeta^{\alpha}_i \partial_{u^{\alpha}_i} +\zeta^{\alpha}_{i_1 i_2}
   \partial_{ u^{\alpha}_{i_1 i_2}}+\ldots\, ,\\
 & \zeta^{\alpha}_i=D_i({\varkappa}^\alpha)+\xi^j u^{\alpha}_{ij}\,,
 \quad\zeta^{\alpha}_{i_1 i_2}= D_{i_1} D_{i_2}({\varkappa}^\alpha)
     + \xi^{j} u^{\alpha}_{j i_1 i_2}\,, \quad
 {\varkappa}^\alpha = \eta^{\alpha} - \xi^{i} u_{i}^{\alpha}\,,\\
 \end{aligned}   \end{equation}
with coordinates $ \xi^{i}([z,u]) $, $\eta^{\alpha}([z,u]) $,
$\zeta^{\alpha}_i([z, u]) \ldots$, being differential functions
from the space $\cal{A}$. A set of all Lie-B\"acklund  operators
forms an infinite-dimensional Lie algebra $L_{\cal B}$, and any
operator of a form $X_{\ast} = \xi^{i} D_{i}$ is the
Lie-B\"acklund operator for any differential function $
\xi^{i}([z,u])$; the set $L_{\ast}$ of operators $X_{\ast}$
forms an ideal in $L_{\cal B}$. This property allows to
introduce a notion of equivalence of two Lie-B\"acklund
operators $X_1\,, \ X_2 \in L_{\cal {B}}$ if $X_1-X_2 \in
L_{\cal {\ast}}$ (written as $X_1 \thicksim X_2 $). In
particular, any Lie-B\"acklund operator $X\in L_{\cal{B}}$ is
equivalent to the operator (\ref {lb-oper}) with $\xi^{i}=0$,
 \begin{equation} \label{lboper}
 X \thicksim Y = X- \xi^{i} D_{i} = {\varkappa}^\alpha
 \partial_{u^{\alpha}}\,, \quad
 {\varkappa}^\alpha \equiv \eta^{\alpha} - \xi^{i} u_{i}^{\alpha}  \,.
 \end{equation}
The operator $Y $ is known as \textit{the canonical
representation} of $X$, and in notation (\ref{lboper}) we imply
the prolongation of action of the operator on all higher-order
derivatives according to formulas (\ref{lb-oper}). It is
essential, that in the group of infinitesimal transformations
$\cal{G}$ with operator (\ref {lboper}) and the parameter $a$
only dependent variables $u^{\alpha}$ are changed,
 \begin{equation}\label{group}
 u^{\prime \, \alpha} = u^\alpha  + a {\varkappa}^\alpha
  + O(a^2)\,, \quad  z^{\prime \, i} = z^i\, ,
 \end{equation}
while independent variables $z^i$ remain unchanged. This
property has allowed to formulate the concept of symmetry groups
of IDE of form (\ref{rgmnonloc}) as a local group of
transformations $\cal{G}$ with operator (\ref{lboper}), at which
the form of the function $F_{\sigma}$ remains unchanged for any
value of a group parameter $a$. Differentiation of the
appropriate invariance condition, which has been written down
for the function $F_{\sigma}$ dependent on the transformed
dependent variable $u^{\prime \, \alpha}$, with respect to the
group parameter $a$ and transition to a limit $a \to 0 $ gives
determining equations (DetEq). In difference from a case of
basic DE, these DetEq generally also are non-local.
\par

With the help of the canonical operator $Y$, the invariance
criterion for Eq.~(\ref{rgmnonloc}) with respect to the admitted
group can be written down in an infinitesimal form
 \begin{equation}
 {\left. Y F_{\sigma}
 \rule[-7pt]{0pt}{19pt}\right|}_{[F_{\sigma}]}=0\,,
 \quad \sigma = 1 + s\,, \ldots , \, q+s\,,
 \quad  \mbox{where} \quad Y\equiv \int\mbox{d} z
 \, {\varkappa} \left( z \right)\, \frac{{\delta}}
 {{\delta}u\left( z \right)}\,.
 \label{inv-cri} \end{equation}
Meaning generalization of action of the canonical group operator
not only on differential functions, but also on
\textit{functionals}, here in definition of $Y$ variational
differentiation \cite{kov-pjetf-92} is used. For integral
functionals (\ref{nonlocalvar}) a derivative of ${\delta} J /
{\delta} u\left (z\right) $ with respect to a function $u$ is
defined via the principal (linear) part of an increment of a
functional as a limit (if it exists) (see
{\cite{bus-bk-80}})\footnote{Definition of a variational
derivative for arbitrary functional see in \cite{vlt-bk-82}}:
 \[
 \frac{{\delta} J
 \left[u\right]}{{\delta}u\left(z\right)} =
 \lim_{\epsilon\to 0} \frac{J\left[u+{\delta}u_{\epsilon}\right] -
 J\left[u\right]} {\int_{\Delta}\mbox{d}\tau\,
 {\delta}u_{\epsilon}\left(\tau\right)}\,;\quad z \in
 \left({\tau_1},\,{\tau_2}\right)\,.
 \]
Here an infinitesimal variation $\delta u_{\epsilon} (z) \ge 0 $
is a continuously differentiable function given on fixed interval
$ {\Delta} = \left[{\tau_1}, \, {\tau_2} \right]$ which differ
from zero only in ${\epsilon}$-vicinity of a point $z$, and the
norm ${\|{\delta}u_{\epsilon} \|}_{C^1} \,\to 0$ at ${\epsilon}\,
\to 0$.\par

It is checked up by direct calculation that the action of the
operator $Y$ on any differential function and its derivatives, for
example $u, u_z, \, \ldots$ gives a usual result: $Y u =\varkappa,
Y u_z = D_z (\varkappa) $ etc. Hence, if $F_{\sigma} =0$ is a
usual DE, formulas (\ref{inv-cri}) result to local DetEq, and in a
case when $F_{\sigma} =0 \,$ has a form of a system of IDE,
formulas (\ref{inv-cri}) can be considered as \textit{non-local}
DetEq, dependent both upon local and non-local variables.\par

Treating local and non-local variables in DetEq as independent
variables allows to divide these equations into local and
non-local. A procedure of a solution of local DetEq is carried out
in a standard way, using Lie algorithm based on splitting of a
system of overdetermined equations with respect to local variables
and their derivatives. In result expressions for coordinates of
the group operator are found, determining so-called group of
\textit{intermediate} symmetries~\cite{kov-pjetf-92} which are
used further at the analysis of non-local DetEq. Procedure of the
solution of non-local DetEq is carried out similarly, by
substituting coordinates of group operator of intermediate
symmetry found in non-local DetEq and splitting them with the help
of variational differentiation. Hence, construction of symmetries
for non-local equations also becomes an algorithmic procedure,
which can be presented as a sequence of the following operations:
\begin {itemize}
 \item [a)] definition of a set of local group variables,
 \item [b)] construction of DetEq on a basis of infinitesimal
           invariance criterion, \\
           which uses a generalization of the definition of the canonical
           operator,
 \item [c)] separation of DetEq into local and non-local,
 \item [d)] solution of local DetEq with use of the standard Lie algorithm,
 \item [e)] solution non-local DetEq with the help of the
            operation of the variational differentiation.
\end {itemize}
These operations are generalization of the second step of
algorithm on a case, when $\cal{RM}$ is the integral or the
integro-differential manifold.
\medskip

In summary we describe an operation of prolongation of a symmetry
group on non-local variables. To execute standard (in the
classical group analysis) operation of prolongation of the
operator of Lie point group on the non-local variable defined, for
example, by integral relationship (\ref{nonlocalvar}), first write
down this operator in the canonical form, $Y$, and then formally
prolong it on non-local variable $J$
 \begin{equation} \label{prolong-can}
 Y + {\varkappa}^J \partial_{J} \equiv {\varkappa} \partial_u
                      + {\varkappa}^J \partial_{J} \,.
 \end{equation}
The integral relation between ${\varkappa}$ and ${\varkappa}^J$
is obtained by applying operator (\ref{prolong-can}) to
Eq.~(\ref{nonlocalvar}) which was introduced as a definition of
the variable $J$. Substituting explicit expressions for the
coordinate ${\varkappa}$ of the operator $Y$ and calculating the
resulting integrals we obtain the required coordinate
${\varkappa}^J$ of the prolonged operator,
 \begin{equation}\label{canonic3}
 \varkappa^J = \int \frac{\delta J(u)}{\delta u (z)}\, \,
 \varkappa(z)\,    {\rm{d}} z \equiv
 \int \frac{\delta {\cal F}(u(z^{\prime}))}{\delta u (z)} \,\,
 \varkappa(z)\, {\rm{d}} z \, {\rm{d}} z^{\prime} =
  \int {\cal F}_u \,\, \varkappa(z)\,  {\rm{d}} z \,.
 \end{equation}
Here for brevity only one argument of a generator's coordinate is
specified, namely the argument upon which the integration is
fulfilled. \par

The actions described in this paragraph resulting in operators
of the admitted group in non-canonical (\ref{lb-oper}) or in
canonical (\ref{lboper}) representation, make essence of the
second step of RG algorithm.
\bigskip

\subsection{Restriction a group on a solution}{\label{RESTRICTION}}

The group found on the second step, $\cal{G}$, which is
determined by operators (\ref{lb-oper}) and (\ref{lboper}), is
generally wider, than the renormalization group of interest. The
last is related to the concrete particular solution of a
boundary problem, hence with the goal to get RGS it is necessary
to make \textbf{the third} step \textbf{(III)}, consisting in
\textit{restriction} of the group $\cal{G}$ on a manifold, set
by this particular solution. From the mathematical point of
view, this procedure consists in checking vanishing conditions
for a linear combination of coordinates ${\varkappa}_j^{\alpha}$
of the canonical operator equivalent to (\ref{lb-oper}), on some
particular solution $U^{\alpha}(z)$ of a boundary problem
 \begin{equation} \label{restrict}
 {\left\{\ \sum\limits_{j} A^j {\varkappa}^{\alpha}_j
  \equiv \sum\limits_{j} A^j \left(\eta^{\alpha}_j-\xi^i_j
 u_i^{\alpha}\right)\ \right\}}_{
 \Big\vert \displaystyle{u^{\alpha}=U^{\alpha}(z)}}=0 \,.
 \end{equation}
The form of the condition set by relation (\ref{restrict}), is
common for any solution of the BVP, but a way of realization of
the restriction procedure of a group in each separate case is
different. Usually a particular approximate solution for a
concrete BVP is used. On the general scheme it is specified as the
dotted arrow connecting ``initial object'' --- the approximate
solution for particular BVP -- with that object which arises in a
result of the third step. \par

At calculation of combination (\ref{restrict}) on a concrete
particular solution $U^{\alpha}(z)$ it is transformed from a
system of DE for group invariants to algebraic relationships. We
shall note two consequences of the specified actions. First, the
procedure of restriction results in a set of relations between
various $A^j $ and thus ``links'' coordinates of various group
operators $X_j $, admitted by $\cal{RM}$
(\ref{rgmloc})--(\ref{rgmnonloc}). Second, it eliminates (in part
or completely) an arbitrariness which can arise in values of
coordinates $ \xi^{i} $, $ \eta^{\alpha} $ in a case of the
infinite group $\cal{G}$.
\par

As a rule, the procedure of restriction of the group $\cal{G}$
reduces its dimension. At that the general element
(\ref{lb-oper}) of this new group $\cal{RG}$ after performance
of this procedure is represented by a linear combination of new
generators $R_{i}$ with coordinates
 $\hat\xi^{i}$, $\hat\eta^{\alpha}$,
 \begin{equation}  \label{rgo}
 X\ \Rightarrow \ R = \sum\limits_{j} B^j R_j \,, \quad
 R_{j} = \hat\xi^{i}_{j}\partial_{x^i} + \hat\eta^{\alpha}_{j}
         \partial_{u^{\alpha}}\,,
 \end{equation}
and arbitrary constants $B^j$. \par

The set of operators $R_j$, each containing  the required
solution of a problem in the invariant manifold, defines a group
of transformations, which by analogy with RG for models with DE
we also refer to as \textit{renormgroup}.\par

The statement of the third step of the algorithm given in this
section finishes the description of a procedure of construction
of RGS. In the following section it is shown, how RGS are used
for achievement of an ultimate goal of the RG algorithm, namely
improvement of an approximate solution. \medskip

\subsection{Construction of RG-invariant solution}{\label{SOLUTION}}

Three steps described above completely define the regular
algorithm of construction of RGS, but one more, \textbf{the fourth
(IV)}, final step is necessary. This step consists in use of RGS
operators for finding analytical expressions for new, improved (in
comparison with initial) BVP solutions.
\par

From the mathematical point of view realization of this step
consists in use of \textit{renormgroup invariance}
conditions\footnote{Here we should point on existing the analogy
between this concept and concept of functional self-similarity
\cite{shr-tmf-84}, \cite{kov-tmf-99b} (see also discussion in the
Appendix \ref{ap4}).} which are set by a \textit{joint} system of
the equations (\ref{rgmloc})-(\ref {rgmnonloc}) and vanishing
conditions for a linear combination of coordinates
$\hat{\varkappa}^{\alpha}_{j}$ of the canonical operator
equivalent to (\ref{rgo}),
 \begin{equation}  \label{fsinv}
 \sum\limits_{j} R^j \hat{\varkappa}^{\alpha}_j   \equiv
 \sum\limits_{j} B^j \left(\hat\eta^{\alpha}_j-\hat\xi^i_j
 u_i^{\alpha}\right)\ =0 \,.
 \end{equation}
Necessity of use $\cal{RM}$ while constructing BVP solution is
marked on Figure~\ref{fig1} by the dotted arrow connecting these
two objects.
\par

It is clear, that the form of (\ref{fsinv}) is close to
(\ref{restrict}). However, contrary to a previous step,
differential variables $u^{\alpha}_{i}$ in (\ref{fsinv}) are not
replaced with the approximate expressions for BVP solutions
$U(z)\,$ but are treated as usual dependent variables.\par

In the most widespread case, when the renormgroup appears as a
\textit{one parametric Lie point group}, the invariance
conditions yield \textit{first order partial differential
equations}. Solutions of the connected characteristic equations
give \textit{group invariants} (similar to invariant charges in
QFT) through which the required solution of BVP is expressed.
Generally for arbitrary RGS the RG invariance conditions,
written down for any BVP, do not represent characteristic
equations for the Lie point group operator. They can have more
complex form, for example, being represented as a combination of
partial DE and ordinary DE of a high order. However the common
approach to construction of BVP solution as the RG invariant
solution remains in force.
\par

The description of the fourth step finishes the description of
the regular algorithm of construction of RGS for models with
IDE. It should be noted that two last steps are in the same root
as for models with DE. The following section contains a number
of concrete examples showing the ability of RGS algorithm.

\section{Construction of RGS for integral models}

\subsection{The example with functionals of the solution of Hopf
            equation}

Let's proceed with a simple illustrative example, that we have
discussed earlier at the beginning of the paper, i.e. an initial
problem for Hopf equation (\ref{hopf}). We have shown that the
solution of the BVP (\ref{hopf}) can be constructed using any
generator of the RG algebra (\ref{rg-hopf}). Let we are
interested not in the solution in all space, but only in some
solution characteristic at a specific point, say, a value of its
first derivative at a point $x=0$, which formally can be
introduced by a linear functional of $u$,
 \begin{equation} \label{functional}
 u_x(z,0)\equiv u^0_x = - \int\limits_{-\infty}^{+\infty} \textrm{d}x \,
 \delta^{\prime} (x) u(t,x)\,.
 \end{equation}
The dependence of $u^0_x$ upon $z$ can be easily restored using
the prolongation of the linear combination of RG generators
(\ref{rg-hopf}) on the solution functional (\ref{functional}).
We use again the last generator from the list (\ref{rg-hopf}) in
its simplest form with $\epsilon \psi^4 = 1$. Then we write down
this generator in the canonical form and calculate its
prolongation using formulas (\ref{prolong-can}) and
(\ref{canonic3}). Restricting the RG operator obtained after
such prolongation to the space of group variables $\{z,\epsilon,
u^0_x \}$, we get the RG generator for the solution functional
(\ref{functional}). For concreteness we choose $U=-x$ and get
the following generator
 \begin{equation} \label{rg-hopf-example}
  R_4 = \partial_{\epsilon }  -z (u^0_x)^2 \partial_{u^0_x}   \,.
  \end{equation}
The initial condition for the $u^0_x$ at $z=0$ is known,
$u^0_x(z=0)=-1$, hence the use of the invariant of this
generator, ${J}^0 = \epsilon z - 1 / u^0_x = 1 $ restores the
desired dependence $u^0_x=-1/\left( 1 - \epsilon z \right)$,
which is valid from the point $z=0$ up to the singularity point
$z_{sing}=1/\epsilon$. We note that this result is obtained
\textit{without construction of BVP solution}, using only the
appropriate RGS. On the first sight the considered methodical
example and the construction carried out looks clumsily enough
and it is much easier to proceed from the trivial solution
(\ref{solution-hopf}). But in more complex situations the
explicit form of the solution is frequently unknown, whereas it
is possible to construct RGS. In two subsequent sections of the
paper, {\ref{optics}} and {\ref {vlas-maxwell}},  we give
examples that serve as an illustration of this statement.
\par

\subsection{Nonlinear optics: development of a singularity on the  \\
          laser beam axis}{\label{optics}}

We proceed our illustrations with analyzing the BVP for a system
of two first order partial DE, which include Hopf equation as a
particular case and are widely used in gas dynamics, optics and
plasma physics:
 \begin{equation}
 v_z + vv_x - \alpha I_x=0\,, \quad I_z + v I_x + Iv_x =\nu Iv / x \,.
 \label{nl-optics} \end{equation}
For concreteness below the terminology of nonlinear optics is
used where (\ref{nl-optics}) are known as nonlinear geometrical
optics equations (see, for example, the review
\cite{ahm-ufn-67}) and are utilized to describe the evolution of
a laser beam in a nonlinear medium. In this case $I(z,x)$ stands
for the beam intensity and $v(z,x)$ is the derivative of a beam
eikonal with respect to transverse coordinate, $ \alpha $ is a
factor of nonlinear refraction, $z$ and $x$ are coordinates in a
direction of a beam propagation and in a perpendicular
direction, respectively; $ \nu=1 \,$ and $\nu = 0$ for
cylindrical and for plane case, respectively.
\par

The nonlinear medium occupies a half-space $z\geq 0$ and
boundary conditions for equations (\ref{nl-optics}) are set at
$z=0$,
 \begin {equation}
 v(0,x) = 0 \, \quad I(0,x) = {\cal I}(x) \,.
 \label{nl-optics-boundary}
 \end {equation}
The specific choice of the zero value of an eikonal  derivative
on the boundary corresponds to a collimated beam with the
distribution of the beam intensity upon a transverse coordinate
$x$, characterized by the function ${\cal I}(x)$ with the
maximal value equal to unity.
\par

Various analytical approaches (see the papers
\cite{ahm-ufn-67,vla-bk-97,ber-phr-98}) do not provide us with a
universal method for finding solutions to BVP
(\ref{nl-optics})-(\ref{nl-optics-boundary}), suitable for any
geometry of a problem and any boundary data, while the use of RG
algorithm offers new opportunities (see, for example,
\cite{kov-jnopm-97,ksh-phr-01}). In particular, it allows to
prolong the solution, known in a small vicinity of the boundary
$z \simeq 0$ of a nonlinear medium, up to a vicinity of a
solution singularity, which occurrence represents the most
attracting physical effect. The singularity is formed on a beam
axis hence the behaviour of $I(x)$ and $v(x)$ at $x \to 0$ is of
particular interest.
\par

The surprising thing is that this behaviour can be understood
without knowledge of the complete solution via the application
of RG algorithm to two functionals of the BVP solution
(\ref{nl-optics})-(\ref{nl-optics-boundary}), namely, to the
intensity of a laser beam $I^{0}(z) \equiv I(z,0)$ and to the
second derivative of the eikonal ${W}^{0}(z)\equiv v_{x}(z,0)$,
calculated on an axis of a beam and related to this solution by
formal relationships
 \begin{equation}\label{intensity-eikonal}
 I^{0}(z) = \int \textrm{d}x \, \delta (x) I(z,x)\,, \quad
 {W}^{0}(z) = \int \textrm{d}x \, \delta (x) v_x(z,x)\,.
 \end{equation}
Boundary conditions for these functionals with the account of
(\ref {nl-optics-boundary}) are given as
 \begin{equation}  \label{boundary-func}
  I^0(0)=1, \quad {W}^0(0)=0 \,.
  \end{equation}
In spite of the fact that these conditions do not contain data
on the dependence of a beam intensity upon the coordinate $x$,
such information is included in the RGS operator which explicit
form is defined by the profile of the beam intensity $ {\cal
I}(x)$ at $z=0$. We present two examples, corresponding to
cylindrical and plane (``slit'') laser beams with various ${\cal
I}(x)$.
\par

For a cylindrical beam ($\nu =1$) with the parabolic intensity
distribution ${\cal I}(x)=1-x^2$, the RGS generator has the form
\cite{ksh-phr-01}
\begin{equation}\label{gen-beam-par}
 R^{par} = \left( 1-2\alpha z^2 \right) \partial_{z}
 -2\alpha zx \partial_{x} - 2\alpha \left( x-vz \right)
 \partial_{v} + 4\alpha I z \partial_{I} \,.
\end{equation}
To determine the dependence of $I^0$ and $W^0$ on the coordinate
$z$ we prolong (\ref{gen-beam-par}) on non-local variables
(solution functionals) $I^0$ and $W^0$ that gives the following
generator in the reduced space of variables $\{z, I^0, W^0 \}$,
 \begin{equation}\label{max-beam-par}
 R_4 = \left( 1-2\alpha z^2 \right) \partial_{z}
          + 4\alpha I^0 z \partial_{I^0}
          - 2\alpha (1- 2 z {W}^0 )\partial_{{W}^0} \,.
 \end{equation}
The use of two invariants of generator (\ref{max-beam-par}),
$J_1=(1-2\alpha z^2) I^{0}$ and $J_2=W^0(1-2\alpha z^2)+2\alpha
z$ with evident equalities $J_1=1 $ and $J_2=0$, which follow at
the account of boundary conditions (\ref{boundary-func}),
immediately gives expressions
 \begin{equation}
 I^0 = \frac{1}{1-2\alpha z^2 }\,, \quad
 {W}^0 = - \frac{2 \alpha z}{1-2\alpha z^2 } \,.
 \label{int-beam-par}
 \end{equation}
These formulas describe spatial dependence of variables $I^0(z)$
and ${W}^0(z)$, starting from a boundary of a nonlinear medium
$z=0$ up to the point $z_{sing}=1/\sqrt{2 \alpha}$, where the
solution singularity occur, i.e. where the beam intensity and
the eikonal derivative turns to infinity; beyond this point
there is an area of rays intersection, where equations
(\ref{nl-optics}) can not be applied. Expressions
(\ref{int-beam-par}) also follow from formulas, obtained earlier
\cite{ahm-jetf-68} without use of RG algorithm. However, the RGS
algorithm here presents an elegant way of obtaining these
formulas without calculating the complete solution to BVP.
\par

\begin {figure}[!t]
 \includegraphics [width=0.98\textwidth] {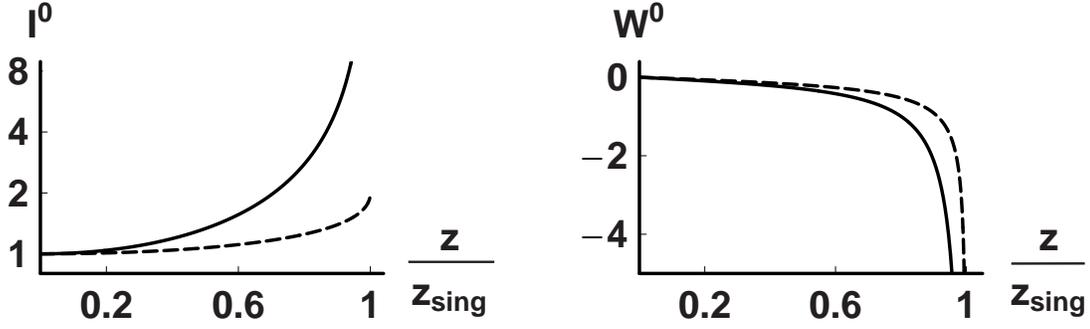}
 \caption {Dependencies of the intensity of a laser beam (at the left) and
 the second derivative of its eikonal (on the right) on the beam axis
 $x=0 $ at various distance $z/z_{sing}$ from the boundary of a nonlinear
 medium $z=0$, plotted with the use of formulas for
 the cylindrical (\ref{int-beam-par}) (block curves) and plane
 (\ref{sol-beam}) (dotted curves) geometry.}
 \label {fig2}
 \end {figure}

Curves of typical dependencies of variables $I^0(z)$ and
${W}^0(z)$ upon the dimensionless coordinate $z/z_{sing}$ at
$\alpha=0.1$ are given on Figure~\ref{fig2}. The change of the
parameter $\alpha$ does not change a type of the curve for the
intensity $I^0 $, whilst values of $W^0$ on the right panel vary
proportionally to $\sqrt{\alpha}$.
Block curves correspond to formulas (\ref{int-beam-par}), i.e. to
parabolic distribution of intensity of a cylindrical beam at the
medium boundary; dotted curves refer to plane geometry of the
beam, considered below.
\par

The procedure of prolongation of the operator on non-local
variables uses a canonical form of RG generators (Lie-B\"acklund
operators) and is suitable also in that case when this generator
is given by a higher-order Lie-B\"acklund symmetry. Such case is
realized for a plane laser beam with ``soliton'' profile of the
intensity distribution at the boundary,
${\cal{I}}(x)=\cosh^{-2}(x)$, when the appropriate RGS generator
has rather cumbersome form
 \begin{equation}  \label{can-gen1-nv}
 \begin{aligned}
 & R^{sol} = \Big{\{} \frac{I}{(Iv_x^2 + \alpha I_x^2)^{2}}
   \left[ \left( \frac{1}{2} \left( Iv_x^2 - \alpha I_x^2 \right)
   \left(v^2 + 4\alpha (1-I)\right) + 4 \alpha v I I_x v_x  \right) v_{xx}
    \right. \\ & \left.
 + \left( 2 \alpha v \left( \alpha I_x^2  - I v_x^2 \right)
 + \alpha v_x I_x \left( v^2 + 4 \alpha (1-I) \right) \right)
            \left( I_{xx}  - \frac{I_x^2}{2I} \right)  \right]
  - v (1-tv_x)  - \alpha t I_x \Big{\}} \partial_v  \\ &
  + \Big{\{}  \frac{I}{(Iv_x^2+\alpha I_x^2)^{2}}
  \Big{[} \left( \frac{1}{2} \left( Iv_x^2  - \alpha I_x^2 \right)
   \left(v^2+4\alpha (1 - I)\right) +  4\alpha v I v_x I_x  \right)
 \left( I_{xx} - \frac{I_x^2}{2I} \right)\\ &
 \vphantom{\frac{a}{b}}-\left[2v\left(\alpha I_x^2- I v_x^2 \right)
   + v_x I_x \left( v^2 + 4 \alpha (1-I) \right) \right] I v_{xx}
  + \frac{1}{4I} (Iv_x^2+\alpha I_x^2)\left[ 4 \alpha I_x^2  \right.  \\ &
 \vphantom{\frac{a}{b}} \left.  + (I_x v-2I v_x)^2 \right] \Big{]}
  - I ( 2- tv_x ) + t v I_x \Big{\}} \partial_I \,. \\
 \end{aligned}
 \end{equation}
Prolongation of (\ref{can-gen1-nv}) on non-local variables
(\ref{intensity-eikonal}) gives the more simple operator in space
of functionals
 \begin{equation}
 \begin{aligned}
 R_5 & \equiv \varkappa^{I^0} \partial_{I^0} +
     \varkappa^{{W}^0} \partial_{{W}^0} =
    \left( 4-5I^0 -zI^0_z + 2 ( I^0 -1) \frac{I^0 I^0_{zz}}{(I^0_z)^2}
    \right) \partial_{I^0} \\ &
  + \left(  \frac{I^0_z}{I^0} + z \frac{I^0_{zz}}{I^0} - z \left(
  \frac{I^0_z}{I^0} \right)^2 - 2(I^0-1)
  \left[
   \frac{I^0_{zzz}}{(I^0_z)^2} + 2 \frac{I^0_z}{(I^0)^2}
   - 2 \frac{(I^0_{zz})^2}{(I^0_z)^3} \right] \right) \partial_{{W}^0}
 \,.   \\
 \end{aligned}
 \label{rgs-pr-sol}
 \end{equation}
While obtaining this formula we used the relation between
derivatives of functions $I$ and $v_x$ with respect to spatial
variables on the beam axis~(at $x=0$), which follows from the
initial equations,
 \begin{equation}
 \begin{aligned} &
  v_{xxx}(z,0) = \frac{1}{\alpha I^0}
  \left[ \frac{I^0_{zzz}}{I^0} + 10 \left( \frac{I^0_z}{I^0} \right)^3
    - 8 \frac{I^0_{zz} I^0_{z}}{(I^0)^2} \right]\,, \\ &
    v_x(z,0) = \frac{I_z^0}{I^0}\,, \quad I_{xx}(z,0) = \frac{1}{\alpha }
  \left[  2 \left( \frac{I^0_z}{I^0} \right)^2
    - \frac{I^0_{zz}}{I^0} \right]  \,.   \\
 \end{aligned}
 \label{spatial-der}
 \end{equation}
The beam intensity and its second eikonal derivative on an axis
are defined from RG invariance condition (\ref{fsinv}), which is
equivalent to vanishing of coordinates $\varkappa^{I^0}$ and
$\varkappa^{{W}^0}$ of the operator (\ref{rgs-pr-sol}). This
condition gives two ODE of the second and the third order
respectively. Solving at first ODE of the second order with the
initial conditions (\ref{boundary-func}) and the additional
condition on the first derivative, $(I^0_z/ \sqrt{I^0-1})_{\vert
z\to 0}=2\sqrt{\alpha}$ which follows from the relation
(\ref{spatial-der}) at $z=0$, we obtain in the implicit form the
law of variation of $I^0$ and ${W}^0$ (compare with formulas
(\ref{int-beam-par}) for a parabolic beam),
 \begin{equation}\label{sol-beam}               
 z = \frac{\sqrt{I^0-1}}{\sqrt{\alpha} I^0 }\,, \quad
 {W}^0 = - \frac{2 \alpha z I^0}{1-2\alpha z^2 I^0} \,.
 \end{equation}
These formulas are valid from the boundary of a nonlinear medium
$z=0$ up to a point where the solution singularity occurs. The
coordinate of the solution singularity is found in view of the
fact that the derivative of ${W}^0$ turns to infinity at this
point, that gives $z_{sing} =1/2\sqrt{\alpha}$, and the value of
the intensity $I^0$ in this point is equal to two. The solution
of the remaining ODE of the third order gives the same result.
Dependencies of $I^0 $ and $W^0 $ upon the dimensionless
coordinate $z/z_{sing}$ at $ \alpha = 0.1 $ are plotted on
Figures \ref{fig2} by dotted curves. Without prolongation of RGS
on non-local variables the result (\ref{sol-beam}) with the use
of RG algorithm was obtained earlier in \cite{ksh-phr-01},
though in a more complicated way.
\par

Summarizing this paragraph, we note, that universality of a
procedures of prolongation of RG generators presented either as
point group operators, or Lie-B\"acklund group operators has
allowed to describe from uniform positions an occurrence of a
singularity of the BVP solution to
(\ref{nl-optics})-(\ref{nl-optics-boundary}), using for this
purpose the reduced description in terms of solution
functionals.

\subsection{RGS for solution functionals of plasma kinetic equations
\label{vlas-maxwell}}

In the paragraph we continue with demonstrations of
potentialities of RGS algorithm for nonlocal models. In contrast
to the previous paragraphs, where we described the prolongation
of the RGS generators on solution functionals for differential
models, here we consider the case when integral relations form a
skeleton of a problem. A vivid example is the model that is used
in the plasma kinetic theory and define the evolution of a
collisionless inhomogeneous plasma. \par

The macroscopic state of plasma particles is governed by
distribution functions $f^{\alpha}$ (for each species of plasma
particles with mass $m_\alpha$ and charge $e_\alpha$), that
dependent on time $t$, a coordinate $x$ of a particle, and its
velocity $v$ (for simplicity we consider the one-dimensional
plane geometry). Evolution of distribution functions is
described by Vlasov kinetic equations {\cite{vla-jetf-38}},
 \begin{equation}
  f^\alpha_t  + v  f_x^\alpha + (e_\alpha/m_\alpha)
  E(t,x)  f_v^\alpha  = 0   \label{planevlasov}
 \end{equation}
supplemented by Poisson and Maxwell equations for the electric
field $E$,
\begin{gather}
  {E}_x - 4 \pi \sum\limits_{\alpha} \int \textrm{d} v
          e_{\alpha} f^{\alpha} = 0 \,, \quad
     {E}_t + 4 \pi \sum\limits_{\alpha} \int \textrm{d} v
     v e_{\alpha} f^{\alpha} = 0 \,.
     \label{planefield}
\end{gather}
The joint system of equations (\ref{planevlasov}) and the first
equation in (\ref{planefield}) is often referred to as
Vlasov-Poisson (VP) equations. We are interested in a solution
to the Cauchy problem to equations (\ref{planevlasov}) with the
initial conditions that correspond to the electron and ion
distribution functions specified at $t=0$
 \begin{equation}\label{initial-cond}
 f^\alpha\big{|}_{ \, {t=0}} = f^\alpha_{0}(x,v)\,.
 \end{equation}
The admitted symmetry group for (\ref{planevlasov}),
(\ref{planefield}), calculated by the method prescribed in
section \ref{GROUP}, is given by time and space translations,
Galilean boosts and two generators of dilations
\cite{kov-de-93a}. VP equations seem to be the simplest
one-dimensional mathematical model, which is commonly used to
describe the evolution of inhomogeneous plasma, e.g., the
expansion of a plasma slab. Even so analytical methods fail to
create the spatially symmetric solution of
(\ref{planevlasov})--(\ref{planefield}) for the distribution
functions with initial zero mean velocity. Thus, with the goal
to find physically reasonable solution we are forced to simplify
the basic system of VP equations.
\par

One possible way to simplify the system (\ref{planevlasov}),
(\ref{planefield}) is to study dynamics of plasma expansion in
quasi-neutral approximation \cite{dor-prl-98, kov-jetfl-01},
suitable for a description of plasma flows with characteristic
scale of density variation large as compared with Debye length
for plasma particles. It means that one can neglect the field
terms in Poisson and Maxwell equations (\ref{planefield}) and
consider the total charge and current densities equal to zero.
Hence, particle distribution functions $f^\alpha(t,x,v)$ for the
electrons ($\alpha = e$) and ions ($\alpha = 1,\,2\, ...$) obey
kinetic equations (\ref{planevlasov}) and are assumed to satisfy
the non-local quasi-neutrality conditions,
 \begin{equation}
 \int \textrm{d} v \, \sum\limits_{\alpha} e_\alpha f^\alpha = 0\,, \quad
 \int \textrm{d} v \,v \, \sum\limits_{\alpha} e_\alpha f^\alpha  =  0\,,
 \label{quasineutral}
 \end{equation}
while the electric field $E$ is expressed in terms of moments of
distribution functions:
 \begin{equation} \label{quasineutral2}
 E(t,x) = \left( \int \textrm{d}v \,v^2\, \partial_x
  \sum_{\alpha} e_{\alpha}  f^\alpha \right)
   \left( \int \textrm{d} v \sum_{\alpha}
  \frac{e_\alpha^2}{m_\alpha} f^\alpha  \right)^{-1} \,.
 \end{equation}
Analytical study of such yet simplified model represents the
essential difficulties, but due to application of RG algorithm
it is possible not only to construct solution at various initial
particle distribution functions \cite{kov-jetfl-01} but also to
find particles density and energy spectra without calculations
of distribution functions for particles in an explicit form.
\par

To construct RGS we consider a set of local (\ref{planevlasov})
and non-local (\ref{quasineutral}) equations as $ \cal {RM} $,
in which the electric field $E(t,x)$ appears as an unknown
function of the coordinate $x$ and time $t$. The Lie group of
point transformations admitted by this manifold is calculated in
a way similar to that used in section \ref{GROUP}. Here, besides
time and space translations, the Galilean boosts and three
operators of dilations, there arises a new projective group
generator \cite{kov-jetf-02}. Precisely this generator enables
to construct a class of exact solutions to the initial problem
that are of interest, as a linear combination of the generator
of time translations and the projective group generator leaves
the approximate PT solution of the initial value problem
$f^{\alpha} = f_0^{\alpha}(x,v) + O(t)$ invariant at $t\to 0$,
i.e. it is the RGS operator
 \begin{equation}  \label{rg-oper}
 R_6 = (1 + \Omega^2 t^2 ) \partial_t  + \Omega^2 t x \partial_{x}
           + \Omega^2 (x-vt) \partial_{v} \,.
\end{equation}
The generator (\ref{rg-oper}) is the only which selects the
spatially symmetric initial distribution functions with zero
mean velocity. The value $\Omega \, $ can be treated as the
ratio of the ion acoustic velocity to the gradient length $L_0$.
\par

Group invariants of the RG operator (\ref{rg-oper}) are particle
distribution functions $f^\alpha$ and combinations
$J_3=x/\sqrt{1+\Omega^2 t^2}$ and $J_4={{v}}^2 +
\Omega^2({x}-{v}t)^2$. Hence, BVP solutions,  i.e. distribution
functions at any time $t\neq 0$, are expressed with the help of
these invariants in terms of initial distributions
(\ref{initial-cond}),
 \begin{equation} \label{dfneutral}
 f^\alpha = f^\alpha_0(I^{(\alpha)})\,, \
 I^{(\alpha)} = \frac{1}{2} J_4
         + \frac{e_\alpha}{ m_\alpha} \Phi_0(J_3) \,.
 \end{equation}
Here the dependence of $\Phi_0 \,$ upon the invariant $J_3 = x/
\sqrt {1 +\Omega^2 t^2} $ is defined by quasi-neutral conditions
(\ref{quasineutral}), and the electric field $E =-\Phi_x$ is
found with the help of a potential
 \begin{equation} \label{potential}
 \Phi (t,x) = \Phi_0 (J_3) \left( 1 + \Omega^2 t^2 \right)^{-1}\,.
 \end{equation}
A variety of examples, illustrating these formulas for the
plasma slab consisting of different groups of hot and cold
electrons and ions of various species, is found in
\cite{kov-jetf-02}.
\par

Distribution functions (\ref{dfneutral}) give exhaustive
information on the kinetics of plasma bunch expansion. However,
for practical applications rough integral characteristics, such
as partial ion density, $n^q(t,x)$, $(q=1,2,\ldots)$ and ion
energy spectra, $d N_q /d \varepsilon$, might be more useful,
\begin{equation}
 \label{spectrneutral}
 \begin{aligned}  &
  n^q  =  \int\limits_{-\infty}^{\infty}\,{\rm d} v f^q(t,x,v) \, ,
  \quad
  \frac{d N_q}{d \varepsilon} = \frac{1}{m_q v}
  \int\limits_{-\infty}^{\infty} {\rm d} x
  \left(  f^q (t,x,v) + f^q (t,x,-v)\right) \, .
  \end{aligned}
  \end{equation}
In view of the complex dependence upon the invariant
$I^{(\alpha)}$ it is not always possible to carry out direct
integration of a distribution function over velocity in the
analytical form, therefore here the procedure of prolongation of
the operator on solution functionals described in section
\ref{RGM} comes to the aid. The density $n^q(t,x)$ is a linear
functional of $f^q$, hence we prolong the generator $R_6$ on the
solution functional (\ref{spectrneutral}) to get the following
RG generator in the space of variables $\{t,x,n^q\}$,
 \begin{equation} \label{rg-n}
 R_7 = (1 + \Omega^2 t^2 ) \partial_t + \Omega^2 t x \partial_{x}
           - \Omega^2 t n^q \partial_{n^q}\,.
 \end{equation}
Two invariants of this generator, namely $J_{3}$ and $J_{4}^q =
n^q \sqrt{1 +\Omega^2 t^2}$ are related for arbitrary $t \neq 0
$ via their initial values: ${J_{3}}_{\mid t=0} = x^{\prime} $,
$ {J_{4}^q}_{\mid t=0} = {\cal N}_{q} (x^{\prime})$. Therefore,
we get formulas that characterize spatial-temporal distribution
of the density of ions of a given species in terms of the
initial density distribution
\begin{equation}
 \label{density}
  n^{q} = \frac{n_{q 0}}{\sqrt{1 + \Omega^2 t^2}}\, {\cal
  N}_{q}\left( \chi \right) \, , \quad
  {\cal N}_q = \int\limits_{-\infty}^{\infty}{\rm d}v f^q_0 \,, \quad
  \chi =  \frac{x}{\sqrt{1+\Omega^2 t^2} }  \,.
 \end{equation}
The general form of $d N_q /d \varepsilon$ is rather complicated
but its asymptotic behavior at $\Omega t \to \infty$ is
described by the same function ${\cal N}_{q}$,
\begin{equation}
 \label{spectrum}
 \begin{aligned}
 \frac{d N_q}{d \varepsilon} \approx
 \sqrt{\frac{2}{m_q \varepsilon}} \frac{n_{q0}}{\Omega }\,
 {\cal N}_q\left(\varepsilon=\frac{m_q U^2}{2}\right)  \,, \quad
 2 \varepsilon /T_q \gg \left( \Omega t \right)^{-2} \,, \quad
 U=\Omega \chi  \,. \\
 \end{aligned}
 \end{equation}
Relations (\ref{density}) express a well-known FS property
\cite{shr-tmf-84} of a solution: the product $n^{q}\sqrt{1 +
\Omega^2 t^2}$, being one of invariants of the RG generator, is
expressed in terms of some universal function ${\cal N}_{q}$ of
another invariant of the generator (\ref{rg-n}), and a form of
this function is set by initial conditions, i.e. it is defined
in terms of initial distribution functions of particles $f^q_0
$. Such representation has a common enough nature and is a
typical property of BVP solutions obtained with the help of RGS
which is pertinent to name $\Phi$-theorem (on analogy with known
$\Pi$-theorem). We briefly discuss this property in concluding
remarks.
\par

In order to illustrate formulas (\ref{density}) we apply them to
plasma  slab that contains ions of several, say two, types (the
index $q=1,2$ corresponds to heavy and light ions, respectively)
with initial Maxwellian velocity distribution functions, and the
electrons obeying a two-temperature Maxwellian distribution
function with densities and temperatures of the cold and hot
components $n_{c0}$ and $n_{h0}$ ($ n_{c0} + n_{h0}= \sum_{q}Z_q
n_{q0} $) and $T_c$ and $T_h$, respectively. In this case the
density distribution and, hence, the ion energy spectrum is
expressed as
\begin{equation}
 \label{densityimp}
  \begin{aligned} &
  {\cal N}_{q} =  \exp \left[ {{\cal E}}
    \left( \frac{Z_q T_{c}}{T_{q}} \right)
   - \frac{{{U}}^2}{2 v_{T_{q}}^2} \left( 1+\frac{Z_q m_e}{m_q}\right)
    \right] \,, \quad q =1,2 \,, \quad
    v^2_{T q}= \frac{T_q}{m_q} \,, \\
   \end{aligned}
 \end{equation}
where the function ${\cal E}$ is defined in the implicit form
 \begin{equation}
 \label{eps}
 \begin{aligned}
 & \vphantom{\frac{a^a}{b^b}}
 n_{c0}
 = \sum_{q=1,2}Z_q n_{q0}
 \exp\left[ \left( 1+ (Z_q T_{c}/T_{q}) \right){{\cal E}}
 -  ( U^2/2 v_{Tq}^2) \right.  \\ &
  \hphantom{n_{c0}  = \sum_{q=1,2}Z_q n_{q0}  \exp 1+}
  \left.
  \times \left( 1+(Z_q m_e/m_q)
  \right) \right] - n_{h0} \exp\left[  \left( 1- (T_{c}/T_h) \right)
  {{\cal E}} \right]  \,. \\
 \end{aligned}
  \end{equation}
\begin{figure}[!t]
 \begin{center}
\includegraphics[width=0.6\textwidth]{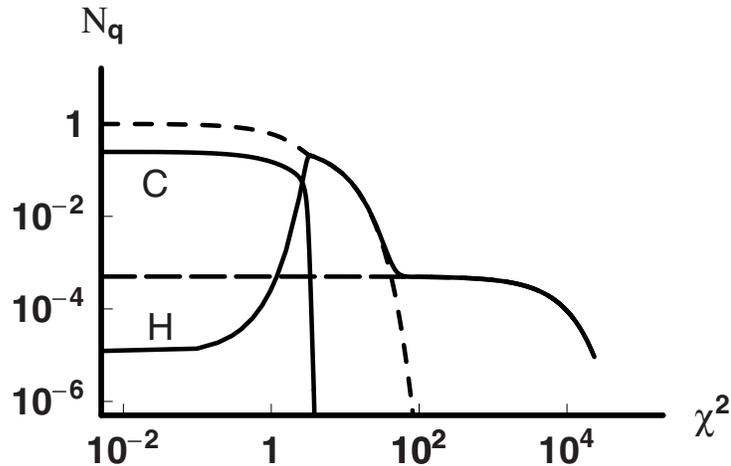}
\caption {The curves describing the dependence of invariants of
         the RGS operator (\ref{rg-n}): the ``universal'' density
         $\rm {N_q}$ of plasma ions -- carbon ions (curves (C)) and
         protons (curves (H)) -- is represented as a function of
         a dimensionless ``coordinate'' $ \chi^2 = (x/L_0)^2/
         (1+\Omega^2 t^2) $. For illustration here
         by dotted curves with short and long strokes the
         dependencies of a dimensionless density for hot and cold
         electrons upon $\chi^2$ are also shown.}
\label{fig3}
 \end{center}
\end{figure}
Figure~\ref{fig3} demonstrates the plots of ${\cal N}_q$ for the
following plasma parameters: $T_{1,2}/T_c=0.1 $; $T_{h}/T_c =
1000 $; $n_{h0}/Z_1n_{10}=5\times 10^{-4}$. Block curves show
dependence of a dimensionless ``universal'' density of plasma
ions $\rm{N_q}=(n_{q0}/n_{c0}){\cal N}_q$, referred to the
maximal density of cold electrons, upon the dimensionless
``coordinate'' $\chi^2=(J_3/L_0)^2 $. Dotted curves present the
distribution of the dimensionless density of cold and hot
electrons (short and long strokes respectively). These curves
demonstrate the high end of the energy spectrum for light ions
$(q=2)$ with a sharp decrease, as well as the spatial separation
of ion species which is of current interest in experiments on
interaction of short laser pulses with thin foil targets (see,
e.g., \cite{kov-fp-04}). Similar results are obtained for more
complex distribution functions \cite{kov-jetf-02} and beyond the
scope of the model used for the one-dimensional expansion, for
example for spherically-symmetric expansion of a plasma bunch
\cite{kov-prl-03}. \par

Summarizing the paragraph \ref{vlas-maxwell} let's note, that
here a modified RG algorithm for calculation of RGS for
non-local systems of the equations, and a procedure of
prolongation of RG operators on solution functionals are used
simultaneously.

\medskip

\section{Conclusion}

The realization of the program that expands opportunities of
application of the RG algorithm to problems of mathematical
physics, is the goal of the paper, which is obviously formulated
in section 1 for non-local problems. The appropriate class covers
now (besides problems on a basis of DE, discussed in section 2 in
an introductory example) the models containing non-local terms,
including integral and inetgro-differential equations. \par

This formulation preserves the former general  scheme of
construction of RG algorithm as four consecutive steps
\cite{ksh-phr-01}. However the form of realization of these
steps significantly varies, that is most brightly shown on first
two stages of the algorithm (see figure \ref{fig1}), related to
construction of non-local RG manifold and a calculation of
admitted symmetry group. Here in view of absence of a regular
computational algorithm (similar to Lie algorithm for DE) while
performing of the second stage of algorithm various realizations
are possible. As an illustration we choose and state in more
detail the variant that is based on use of a canonical operator.

In the following section 4 efficiency of use of procedure of
prolongation the RG operator on non-local variables is shown with
the purpose of the reduced description of the solution in terms of
the integrated characteristic, the solution functional. It is
essential, that the knowledge of a solution in an explicit form
therewith is not required. \par

The examples given here serve as an illustration of the program
formulated in section 3. They show application of new algorithm
to integro-differential systems (expansion of a plasma bunch)
and a procedure of prolongation on solution functionals (Cauchy
problem for Hopf equation, evolution of a laser beam in
nonlinear optics). The example presented in \ref{vlas-maxwell}
is the most valuable, as here the calculation of RG symmetries
for a solution of the non-local system of equations is
supplemented by a procedure of prolongation of the RG operator
obtained on the solution functional, the density of plasma
particles, with the purpose of revealing its variation law. \par

At the formulation and discussion of the analytical form of
results an accent is made on a role of invariants of appropriate
RG operators. The manifested common regularities are considered
in the Appendix.

The results of section 4 testify to universality of a method of
RenormGroup Symmetries. Therefore they allow to look forward to a
further expansion of a class of the problems which can be
investigated with the help of RGS method, and to new objects, for
which the application of RG algorithm yet is not a standard
procedure.\par

Here we mean an infinite systems of looped integro-differential
equations, similar to systems for correlation functions in
statistical physics and to systems of the equations for
generalized Green functions -- propagators and vertex functions --
in the quantum field theory.

\medskip

The work was partially supported by RFBR grant No.05-01-00631,
grant of Scientific School No.2339.2003.2 and ISTC project 2289.


\newpage

\section{Appendix.
 Invariant representation of solution and
 $\Phi$-theorem}{\label{ap4}}

 In the main text of this paper the role of invariants of RG
 transformation in construction of BVP solutions was repeatedly marked.
 Here we consider the relation of representations of BVP
 solution with the concept of functional self-similarity and
 with the well-known principles of the group analysis. \par

Let us remind that BVP solutions which are obtained with the use
of RG algorithm are invariant solutions of RG operators. In
group analysis of differential equations the explicit expression
of solutions through invariants uses the well-known \textbf{
theorem of invariant representation} of a regular (non-singular)
manifold (see. \cite[\S18,\,ch.5]{ovs-bk-82}, and
\cite[Vol.3,\,p.6]{ibr-spr-94}):

 \begin{quote}
 {\small Let a manifold $M\subset \mathbb{R}^N$ admit a group $\cal{G}$.
 Suppose $M\,$ be a nonsingular manifold of a group $\cal{G}$, i.e., an
 infinitesimal operator of group $G$ does not vanish identically on
 $M\,$. Then, $M\,$ can be represented by a system of equations,
 left-hand sides of which are invariants of group ${\cal G}$, i.e.,
  have the form:
 \begin{equation} \label{inv-rm}
 \Phi_k(J_1(z),J_2(z),\ldots,J_{N-1}(z))=0, \ k=1,\ldots,s.
 \end{equation}}
 \end{quote}
Here $ J_1(z),J_2(z),\ldots,J_{N-1}(z), \ z \in \mathbb{R}^N$,
form a basis of invariant of ${\cal G}$. Hence, equations
(\ref{inv-rm}) with arbitrary functions $\Phi_k$ of $N-1$
variables furnish the general form of non-singular invariant
manifold of the group ${\cal G}$. In particular, it gives a
transparent comment of the well-known $\Pi$-theorem
\cite{self-sim-15}.
\par

Turn now to BVP solutions which are obtained with the use of RG
algorithm. For RG invariant solutions there exists a more
general statement as compared with $\Pi$-theorem,
 \begin {quote}
 $ \bm {\, \Phi}$-\textbf{theorem}: \textit{An invariant solution
 of a boundary-value problem can be represented by a system of
 equations of the form (\ref{inv-rm}) written down in terms of
 functional invariants $\, \phi_i \,$ of the problem}.
  \end {quote}
These  $\phi_i\,$ are understood as invariants of appropriate
functional transformations involving not only dependent and
independent variables of the equations, but also parameters of
boundary conditions, that is invariants of renormgroup
transformations.\par

In essence, $\Phi\,$--theorem is an analogue of the
\textit{theorem of invariant representation} with reference to
solutions of BVP having the property of the functional
self-similarity. In this case, one should consider sub-manifold
$\cal{RM}\,$ invariant with respect to renormgroup $\cal{RG}\,$
as a nonsingular manifold.

In a special case, when in (\ref{inv-rm}) $s=1\,,$ and in the
functional invariant $\phi(y,\{a\})\,,$ containing the required
function  $y\,,$ variables are separated, the solution can be
written down in an explicit form close to the representation,
which emerges from the $\Pi$-theorem:
\begin{equation} \label{phi-theorem}
  y=\phi^{-1}_{(1)}(\Phi(\dots), \{a\})\,;  \quad  \quad
 \phi\,= \Phi(\dots,\phi_i,\dots)\,, \quad i=1,\ldots, N-1 \,.
 \end{equation}
Here, the function $\phi^{-1}_{(1)}\,$ is a reverse one to
$\phi\,$ with respect to its first argument. Due to this
solution $y\,$ appears dependent not only on the remaining
functional invariants, $\phi_i\,,$ but also on variables and
parameters, $\{a\}\,,$ entering into the invariant $\phi\,$.
Thus, as well as for power self-similarity, BVP solutions are
not, generally, invariants of RG transformations, but are
expressed  through certain combinations of invariants of RGS
operators.
\par

Expressions (\ref{dfneutral}) for distribution functions of plasma
particles in an expanding bunch serve as an example of BVP
solution, being such invariants.
\par

As the second example, we take a QFT model with two coupling
constants $g\,$ and $h\,.$ Here, invariant quantities, e.g., observed
effective scattering cross--sections $\sigma_{\nu}(s)\,,$ are
expressible in terms of RG invariants --- two invariant coupling
functions $\bar{g}(s/\mu^2;g,h)\,,$  $\bar{h}(s/\mu^2;g,h)\,$ and of
 the ratio $m^2/s\,$ ---  by relations
\begin{equation} \label{sigma-vu}
 \sigma_{\nu}(s)= \Sigma_{\nu}\left(m^2/s,\,\bar{g}\,,
 \,\bar{h}\,\right)\,.
\end{equation}
In turn, functions $\bar{g}\,$ and $\bar{h}\,$ should be found from
system of two functional relations  (see, e.g., eqs.(48.37) in Ref.
 \cite{bsh-bk-80})
\begin{eqnarray}\label{48.37}
G(y/x,~\bar{g}(x,y;g,h),~\bar{h}(x,y;g,h))&=&G(y;g,h)~; \cr
H(y/x,~\bar{g}(x,y;g,h),~\bar{h}(x,y;g,h))&=&H(y;g,h)~,
\end{eqnarray}
containing two arbitrary functions, $G\,$ and $H\,,$ of two arguments.
Due to this, to find each of $\sigma_{\nu}(s)\,,$ one needs to have
explicit expressions for three defining functions
$\,\Sigma_{\nu},\,G\,$ and $H\,.$
\par

Note also that the procedure of numerical defining of the parameters
$g\,$ and $h\,$ from boundary data (in fact, from observed quantities)
involves at least two implicit relations (\ref{sigma-vu}).

At the same time, functionals of functions which determine, according
to formulae (\ref{density}), the density distributions of particles of
expanding plasma, are \textit{not} invariants of the RG operator.
Another example when a BVP solution is constructed with the help of
invariants of RG transformations, but is not such invariant itself,
is submitted by formulae (\ref {int-beam-par}) and (\ref {sol-beam})
for functionals (\ref{intensity-eikonal}) in a problem of a beam
refraction in a nonlinear medium.
Generally, when it is impossible to express the BVP solution in an
explicit form, one should use general formulae (\ref{inv-rm})
instead of the representation (\ref{phi-theorem}).

\end{document}